\documentclass[times,10pt,twocolumn]{article} 

\makeindex
\usepackage{latex8}
\usepackage{amssymb}
\usepackage{graphicx}
\usepackage{xcolor}
\usepackage[utf8x]{inputenc}
\usepackage{url}

\usepackage{wrapfig}

\usepackage{listings}

\usepackage{makeidx}

\usepackage[T1]{fontenc}
\usepackage{tabularx}
\usepackage{array}
\usepackage{multirow}
\usepackage{lscape}
\makeatletter
\def\and{%
  \end{tabular}%
  \hskip 0.1em \@plus.17fil\relax
  \begin{tabular}[t]{c}}
\makeatother

\newcommand{\Crysys}{\texttt{CrySys}}
\newcommand{\Duqu}{\texttt{Duqu}}
\newcommand{\Stuxnet}{\texttt{Stuxnet}}
\newcommand{\driver}{\texttt{nfrd965.sys}}
\newcommand{\netpDLL}{\texttt{NETP191.PNF}}

\newcommand{\netpCONF}{\texttt{netp192.PNF}}

\newcommand{\krn}{\texttt{kernel32.dll}}
\newcommand{\service}{\texttt{services.exe}}
\begin{document}
\title{Analysis and Diversion of Duqu's Driver}
\author{Guillaume Bonfante\\
Universit\'e de Lorraine\\ LORIA \\ 
\and
Jean-Yves Marion\\
Universit\'e de Lorraine\\ LORIA \\ 
\and
Fabrice Sabatier\\
Inria\\
LORIA\\
\and
Aur\'elien Thierry\\
Inria\\
LORIA\\ 
}

\email{firstname.lastname@loria.fr}



\maketitle
\thispagestyle{empty}

\begin{abstract}
The propagation techniques and the payload of \Duqu\ have been closely studied over the past year and it has been said that \Duqu\ shared functionalities with \Stuxnet. We focused on the driver used by \Duqu\ during the infection, our contribution consists in reverse-engineering the driver: we rebuilt its source code and analyzed the mechanisms it uses to execute the payload while avoiding detection. Then we diverted the driver into a defensive version capable of detecting injections in Windows binaries, thus preventing further attacks. We specifically show how \Duqu's modified driver would have detected \Duqu.

\end{abstract}


\section{Introduction}

When it was first discovered in September 2011 by \Crysys\ \cite{AThierry_CrysysDuquStuxnet}, it was said that \Duqu\ was close to \Stuxnet\ in that they shared infection and propagation mechanisms. 
\Duqu\ is an offensive tool used to steal information: Symantec \cite{AThierry_SymantecDuqu2011} identified, amongst other payloads, keyloggers, screen recorders, network monitors and service discovery tools.
It is kept up to date by Command \& Control servers and bundled with an auto-destruction feature typically triggered 36 days after the infection.

The attacks seem to have been successful since the malware was not detected during the operations (some of them lasted a few month) but only post-mortem.
Besides, according to Kasperky \cite{AThierry_KaspDuqu10}, the latest version of \Duqu\ was found in February 2012, long after the malware was first discovered and documented.
Thus \Duqu\ is a stealthy spyware and the attacks were carefully adapted to each target.


At the High Sec Lab, we develop a malware detector based on morphological analysis~\cite{AThierry_BKM08}, which is a technique based on control flow graph comparison. As soon as we put our hands on \Stuxnet\ and \Duqu\ samples, we wanted to know if our approach, knowing \Stuxnet, could lead to the detection of \Duqu. In fact the answer is \emph{yes but}: we recognize \Duqu's main DLL as related to \Stuxnet\ but only if it is decrypted. 
The aim of this paper is to present an automatic extraction procedure to bypass stealth techniques such as those used by \Duqu.
The irony of this work is that \Duqu\ itself served for this purpose. Let us describe in broad terms our contribution.

\paragraph{Infection timeline.}
The infection detected by \Crysys\ used a malicious Microsoft Word document embedding \Duqu.
Firstly it exploits a $0$-day kernel vulnerability (on TrueType fonts \cite{AThierry_CVETrueType}), it installs three components: a driver (\driver), an encrypted DLL (\Duqu's main DLL, \netpDLL) and an encrypted configuration file (\netpCONF).

The second step takes place at the next reboot. The driver monitors processes loaded by the OS and injects \Duqu's DLL into a process specified by the configuration file, typically \service.

During the third and final step, the payload, which is included in the DLL, is activated.

Detecting this malware is challenging because only the driver is unencrypted on the hard drive.
The DLL is encrypted, packed with UPX, and is injected immediately after decryption so it is only stored decrypted in RAM.
Thus detection could take place when the injection is done, which is right after the decryption of the DLL and before it is activated.





Once installed on a first target, \Duqu\ receives orders for further attacks and propagation from Command \& Control servers.
Each new controlled machine can be configured to connect back to the attacker through some routing tunnel, providing access to restricted areas.

\paragraph{Defensive version.}
The defense scenario will be the following. We modify \Duqu's driver so it monitors loaded processes,
stores hashes of their code and detects when an alteration is done.

We began by rebuilding a consistent source code for the driver from the binary, then performed an in-depth analysis of the infection method. Finally we modified the code to build a defensive version capable of monitoring loaded processes and detecting an infection.




\section{Rebuilding the driver's source code}

We knew \Duqu\ shared code with \Stuxnet\ since we showed equivalent parts in the drivers' main DLLs \cite{AThierry_REAT12} and Symantec revealed similarities in the injection technique and some resources. The decompilation of \Stuxnet's driver had already be done by Amr Thabet \cite{AThierry_ThabetDriver} but we decided to decompile \Duqu's driver into C code to expose how it operates and the difference between the drivers of \Duqu\ and \Stuxnet. And actually we found singular techniques for injection and anti-detection mechanisms.




\subsection{IDA's decompiler}
We used IDA's plugin "Hex-Rays Decompiler" \cite{AThierry_IDADecompiler}. It generates C pseudocode from a binary being analyzed in IDA and authorizes modifications of the C code from within the plugin's GUI. 
Unfortunately the code generated by IDA was, in our case, not directly usable: an example of the first lines of a function decompiled by IDA is given on Figure \ref{fig:AThierry_ParsePEInitial}.
Firstly the code cannot be re-compiled because some variable types, function arguments and calling conventions were not recognized by the decompiler.
Secondly the generated code is hardly readable, partly because some structures and librairies were not identified. Let us detail how to circumvent these issues.


In order to rebuild a consistent source code we followed an incremental protocol based on the decompiler's output. We began with commenting the whole decompiled code except for the entrypoint of the driver. We fixed compilation errors and identified missing structures. Once the entrypoint was readable and could be compiled, we added some more code that was previously commented, fixed it, and so on.



\subsection{Structures and types identification}

Figure \ref{fig:AThierry_ParsePEInitial} depicts the first lines of the decompiler's output for one of the driver's function.
A lot of information is missing, for instance most types are described as \texttt{int} so the kind of processing done is unknown.
We will give more details on this example to illustrate what has been done to fix the code generated by the decompiler.
Some of the constants are going to be of great help: for instance we can combine both of them 0xF750F284~XOR~0xF750B7D4~=~0x00004550 =~'PE$\backslash$0$\backslash$0', the text searched is 'PE$\backslash$0$\backslash$0' which is obfuscated with a XOR operation.

At this point we have strong suspicion that this function is used to parse PE files.
A quick dive in Microsoft's Visual C++ documentation reveals the structure PIMAGE\_NT\_HEADERS which first field is \emph{Signature}, whose value is 'PE$\backslash$0$\backslash$0' for Windows binaries. Thus the variable \texttt{v4}, which is tested against 'PE$\backslash$0$\backslash$0', might be of type PIMAGE\_NT\_HEADERS.
So we force this structure and others into IDA's decompiler in place of default \texttt{int} type, IDA then finds fields based on their offsets and updates them in the code. In some cases the structure is specific to the analyzed binary (user defined), it is possible to define them manually in the decompiler plugin.

The beginning of the fixed code is presented on Figure \ref{fig:AThierry_ParsePEFinal}, it is readable by a C developer: one sees that the function checks whether a file is a PE binary. The following part of the code will parse it and fill a user defined structure with some information (entrypoint, sections...). Besides the compiled binary of this version is very similar to the original binary. Doing this kind of analysis on the whole driver lead us to build an understandable and consistent version of the driver's source code.

\begin{figure}

\begin{small}\begin{lstlisting}[language={C}]
signed int __cdecl sub_12F36(int a1,
		     int a2, int a3)
{
  int v4; // eax@3
  unsigned __int16 v5; // cx@4
  int v6; // ecx@7

  v4 = a2 + *(_DWORD *)(a2 + 60);
  if (*(_DWORD *)v4 ^ 0xF750F284
	           != 0xF750B7D4)
    return 1;
\end{lstlisting}\end{small}
\caption{First 10 lines of the ParsePE function as decompiled by IDA\label{fig:AThierry_ParsePEInitial}}
\end{figure}

\begin{figure}
\begin{small}\begin{lstlisting}[language={C}]
NTSTATUS __cdecl ParsePE(
 __out PEDataPtr pPEData, 
 __in PIMAGE_DOS_HEADER BaseAddress,
 __in int flag){
PVOID infosPE;
PIMAGE_DOS_HEADER pDosHeader;
PIMAGE_NT_HEADERS pNtHeader;

pNtHeader = (DWORD)infosPE +
		  infosPE->e_lfanew;
if ((pNtHeader->Signature ^ 0xF750F284)
  != (IMAGE_NT_SIGNATURE ^ 0xF750F284)) 
    return STATUS_WAIT_1; 
\end{lstlisting}\end{small}
\caption{First 10 lines of the fixed ParsePE function\label{fig:AThierry_ParsePEFinal}}
\end{figure}

\section{Functional analysis from the source code}

Now that the source code has been rebuilt, let us detail how it operates.
There are two main phases, the first one consists in setting up the driver: it asks the operating system for notifications when a binary is loaded, and initializes stealth mechanisms.
The second one is triggered when it receives notifications: the driver infects the target binary by injecting \Duqu's DLL into \service, then activates the payload.


\subsection{Initialisation of the driver during boot}
Recall that on Windows, the startup order of drivers is determined by their \texttt{Group} key in the registry. \Duqu's \driver, belonging to the "network" group, is activated before the hardware abstraction layer (HAL) is loaded into memory.


Once started, \driver\ allocates 512 bytes for storing a pointer array of functions shared by various callback routines. Then it decrypts some internal parameters, revealing the name and path of the registry key used for configuring the injection.

%
%
%

If the decryption succeeded, the driver checks its execution mode. If it is in debug or fail-safe mode, the driver halts, otherwise it creates a device named \texttt{\{624409B3-4CEF-41c0-8B81-7634279A41E5\}} and defines a list of control commands that the device can process.

That being done, it registers two callback functions within the kernel's event handler. The first is required by the operating system: it is used to create an access point  ($\backslash$\texttt{Device}$\backslash$\texttt{Gpd0}) and a link ($\backslash$\texttt{DosDevices}$\backslash$\texttt{GpdDev}) to the driver, and attaches the device to a memory stack.
The second function will be called when the driver is initialized or reinitialized, it is inserted in a waiting list of events.

This second function waits until the Windows kernel is totally loaded: it checks if the DLL \texttt{hal.dll} is loaded, if not, the function is once again inserted into the events waiting list (for at most 200 times).
When the system is ready, an access point $\backslash$\texttt{Device}$\backslash$\texttt{Gpd1} is created and linked to a request processing function. 
At that point librairies are available for \Duqu's injection.

\subsubsection{Stealth techniques}
The driver acts like a rootkit because it avoids directly using suspicious system calls. Indeed those system calls might be monitored by an antivirus. Typically the function \texttt{ZwAllocateVirtualMemory} can be used to allocate memory space into any process, for instance to inject code into any target process. Besides, in order to hook the entrypoint of a system binary, \Duqu\ also needs the function \texttt{ZwProtectVirtualMemory} that is deliberately not exported by the kernel. This function modifies the permissions of a memory block and can be used to make a code section writable or execute code in a data section. \Duqu\ was built to find \texttt{ZwProtectVirtualMemory}'s memory address without using imports.

The two functions are implemented in the Windows kernel (in \texttt{Ntoskrnl.exe} or \texttt{ntkrnlpa.exe}, depending on the version).
The driver inspects every module (DLLs and EXE files) loaded by the OS during boot until it finds one of the two target kernel files.

Once the file is found, the driver uses the function \texttt{ParsePE} to examine it closely.
It searches, in that kernel file, a call to \texttt{ZwAllocateVirtualMemory} (whose address is known from the export table) followed by the opcode \texttt{push~104h} and another (near) call to an unknown function. If this pattern, shown in Figure \ref{fig:AThierry_CallZwProtect}, is found, the target address of this last call is considered to be \texttt{ZwProtectVirtualMemory}.
At this point the memory addresses of both functions are known.

\begin{figure*}
\scriptsize
\begin{lstlisting}[language={[x86masm]Assembler}, escapechar=~]
(01) PAGE:004ED1AD                  loc_4ED1AD: [...]                      
(02) PAGE:004ED1BC 50               push    eax             ; BaseAddress
(03) PAGE:004ED1BD 57               push    edi             ; ProcessHandle
(04) PAGE:004ED1BE E8 19 8C F1 FF   ~\textcolor{red}{\texttt{call    DS:ZwAllocateVirtualMemory}}~
(05) PAGE:004ED1C3 3B C3            cmp     eax, ebx
(06) PAGE:004ED1C5 8B 4D FC         mov     ecx, [ebp+BaseAddress]
(07) PAGE:004ED1C8 89 4E 0C         mov     [esi+0Ch], ecx
(08) PAGE:004ED1CB 7C 2E            jl      short loc_4ED1FB
(09) PAGE:004ED1CD 38 5D 0B         cmp     byte ptr [ebp+ProcessHandle+3], bl
(10) PAGE:004ED1D0 74 27            jz      short loc_4ED1F9
(11) PAGE:004ED1D2 8B 45 D0         mov     eax, [ebp+var_30]
(12) PAGE:004ED1D5 89 45 F8         mov     [ebp+ProtectSize], eax
(13) PAGE:004ED1D8 8D 45 F4         lea     eax, [ebp+OldProtect]
(14) PAGE:004ED1DB 50               push    eax             ; OldProtect
(15) PAGE:004ED1DC 68 04 01 00 00   ~\textcolor{red}{\texttt{push    104h}}~
(16) PAGE:004ED1E1 8D 45 F8         lea     eax, [ebp+ProtectSize]
(17) PAGE:004ED1E4 50               push    eax             ; ProtectSize
(18) PAGE:004ED1E5 8D 45 FC         lea     eax, [ebp+BaseAddress]
(19) PAGE:004ED1E8 50               push    eax             ; BaseAddress
(20) PAGE:004ED1E9 57               push    edi             ; ProcessHandle
(21) PAGE:004ED1EA E8 93 96 F1 FF   ~\textcolor{red}{\texttt{call    loc\_406882}}~ ; ZwProtectVirtualMemory
(22) PAGE:004ED1EF 3B C3            cmp     eax, ebx
\end{lstlisting}
\caption{Function calling ZwProtectVirtualMemory.\label{fig:AThierry_CallZwProtect}}
\end{figure*}

\begin{figure}
\begin{center}
 \includegraphics[width=0.48\textwidth]{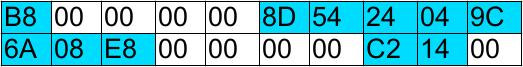}
 \includegraphics[width=0.48\textwidth]{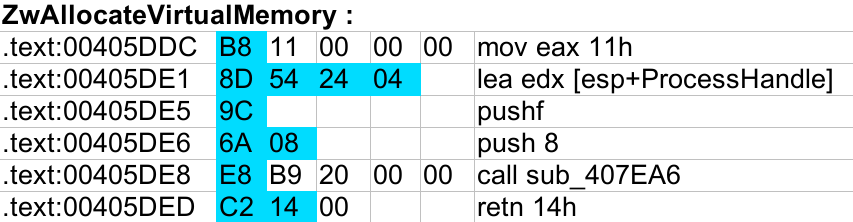}
 \includegraphics[width=0.48\textwidth]{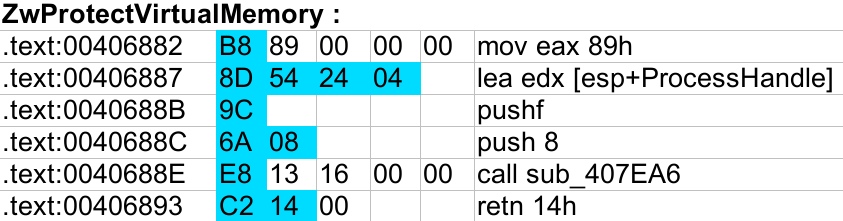}
\end{center}
\caption{Integrity mask applied on ZwAllocateVirtualMemory and \mbox{ZwProtectVirtualMemory}. Values filled in gray are those checked.}
\label{fig:AThierry_matching}
\end{figure}

\paragraph{Avoiding hooks.}
It is firstly checked that the two functions are located inside the kernel's memory addresses and not in user space which would indicate an obvious hook to a monitored function.
Secondly an integrity mask applying a logical \texttt{AND} is applied on the first 32 bytes of both functions. The mask is the same for the two functions and is shown on Figure \ref{fig:AThierry_matching}.
If the functions pass both tests, their addresses are considered valid and are kept for a future stealthy use.

\subsubsection{Initialization of shared memory}
A shared memory space is allocated and used as a link between the driver's  callback functions and the kernel. It is to be filled with, amongst others, infection parameters decrypted from the registry and an import table giving access to \krn\ and kernel functions. This import table will be used by both the executable code that \Duqu\ is going to inject and the payload.

The initialization phase ends by setting up a notification triggered each time a module is loaded into memory (through the system call \texttt{PsSetLoadImageNotifyRoutine}).

%

\subsection{Code injection}
\subsubsection{Processing the first notification}
\paragraph{Before the injection.}
The driver is notified each time a module (DLL or EXE file) is loaded into memory.
Each time, the driver checks if the Windows version is supported, then it tries to locate the mapped module. To do so it uses the process \texttt{id} given as a parameter by the OS. It reads the file's base address from the PEB (Process Environment Block) structure and compares it to the address passed by the operating system. It checks that the configuration file is decrypted in the shared memory and reads the target file field. 
As explained by \Crysys' document \cite{AThierry_CrysysDuquStuxnet}, the target file in that case is \service\ so from now on we will focus on that process and the injection into it.


\paragraph{Payload injection.}



\Duqu's driver is now going to inject malicious code into \service. Thus the payload will be executed by \service\ before \service's legitimate code.

Once \service\ is loaded, the driver determines its entrypoint and allocates memory (with \texttt{ZwAllocateVirtualMemory}) in the \texttt{.data} segment. It injects two \texttt{PE} files with altered headers. Then it restores the missing constants ('\texttt{MZ}', '\texttt{IMAGE\_NT\_SIGNATURE}', '\texttt{IMAGE\_PE\_i386\_MACHINE}, and '\texttt{IMAGE\_PE32\_MAGIC}') of the first injected code.
Finally it proceeds with the addresses relocation and modifies the permissions of \service's entrypoint from \texttt{RX} (\texttt{PAGE\_EXECUTE\_READ}) to \texttt{RWX} (\texttt{PAGE\_EXECUTE\_WRITECOPY}) using \texttt{ZwProtectVirtualMemory}.

\Duqu's \driver\ allocates memory in the \service\ process, its size is 57 bytes plus the size of the decrypted DLL. The payload (the DLL \netpDLL) is decrypted and copied there.
Next a \emph{handler} is opened on the kernel driver (\driver), saved in a shared structure in order to be gathered by the injected code.

\subsubsection{Processing the second notification}
The driver is not only notified when the main module (process \service) is loaded, but also when DLLs linked to that module are loaded. In particular, when \krn\ is loaded, the driver looks for the addresses of 10 functions exported by \texttt{kernel32.dll} that will be used by the payload. Trying to be stealthy, the search consists in comparing the hashed names of 10 functions.
This processing ends with saving the first 12 bytes of the entrypoint assembly code of \service\ and their replacement by a jump on the first injected (and restored) code. The first instructions of the entrypoint are changed into \texttt{mov~eax,@adresseInjection} followed by \texttt{call~eax}.

The process \service\ has been altered and is now ready to launch the payload.

\subsection{Launching the payload}
The operating system finishes the initialization of the \service\ process and proceeds with its execution by passing control to the code at the entrypoint. Actually the system starts the first injected code.

Its first task consists in determining its own absolute memory address (with the instruction sequence \emph{call-pop}) because further processing (read, write, jump) depend on it. During execution the addresses are relocated with respect to the absolute address of the entrypoint.

It then restores the headers of the second injected code so it is a valid \texttt{PE} and fills, within a shared structure, an import table from the 10 functions found in \texttt{kernel32.dll}.
Then it creates a handler on \texttt{ntdll.dll} which is stored in a shared structure. It then jumps to the entrypoint of the second injected code.  

\begin{figure}
\begin{center}
\includegraphics[width=0.4\textwidth]{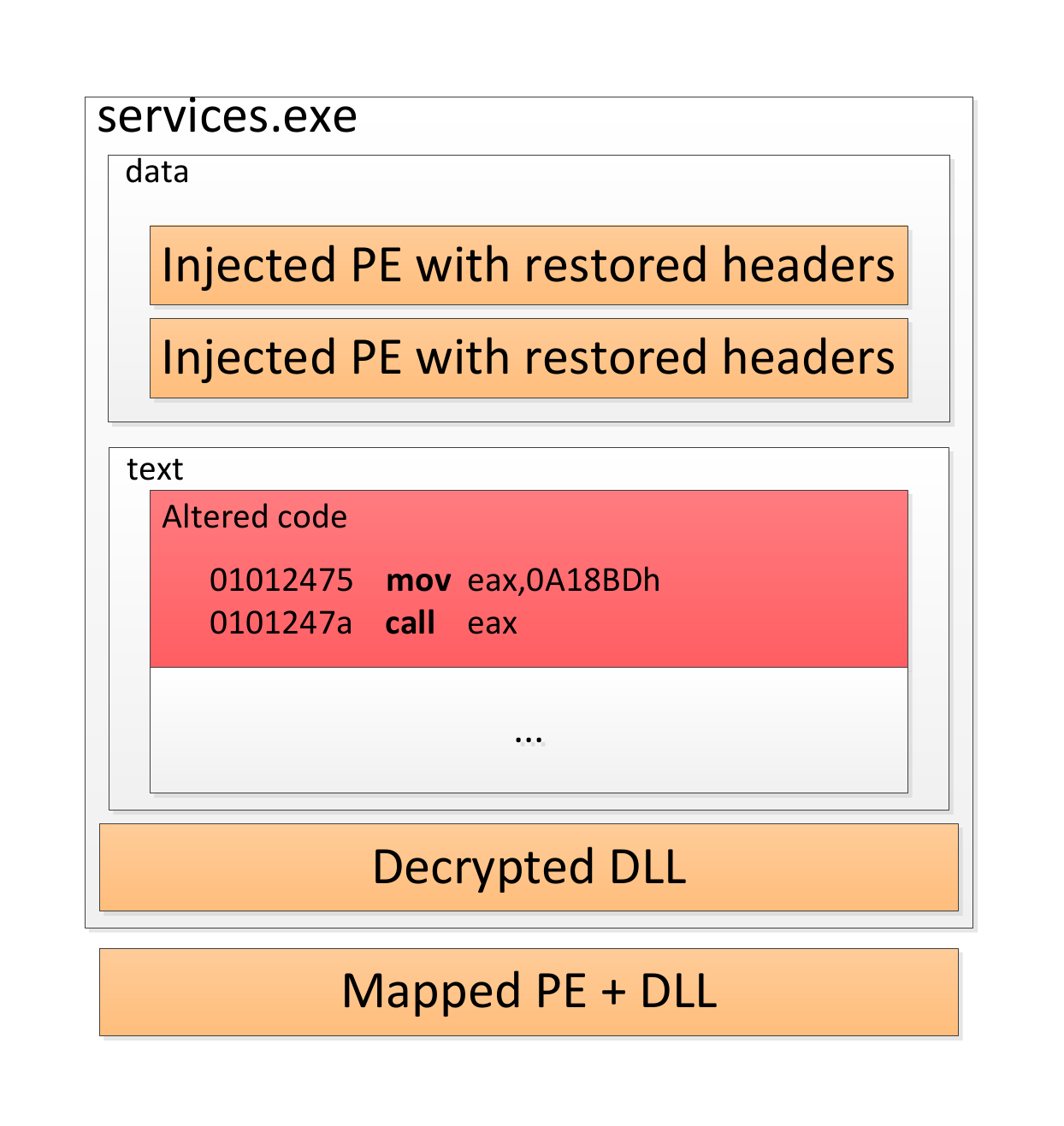} 
\end{center}
\caption{Memory space injected by Duqu into services.exe once the injection is done.}
\label{fig:AThierry_ServiceMem}
\end{figure}

This additional module adds data from its own PE header (address of the module, number of sections, address of the export table) to the shared structure. Finally these informations are used to map the PE into memory manually: it allocates memory space, copies the PE header, maps sections, loads DLLs, creates the import table, relocates the addresses and finally determines the entrypoint.
Then the function relocates \Duqu's main decrypted DLL, \netpDLL, as a DLL linked to the PE just mapped and calls its entrypoint. Figure \ref{fig:AThierry_ServiceMem} sums up the injections done by \Duqu\ into \service\ and the system memory.

The payload contained in the DLL is now in place and executed. Once it is done it sends a request to the driver through the access point $\backslash$\texttt{Device}$\backslash$\begin{small}\texttt{\{624409B3-4CEF-41c0-8B81-7634279A41E5\}}\end{small} so it restores the 12 first bytes located at the entrypoint of \service. A second request is finally sent to restore the access permission of \service's entrypoint.

The injection is now done, control is then passed back to the restored \service.

\section{Turning the driver}
\label{section:utilisation}

In the previous paragraph we described how the \Duqu's DLL is injected in \service. Some of these mechanisms, for instance the notifications once a module is loaded, can be used for defensive purposes.

In a nutshell the modified driver will calculate signatures (checksums) on binaries upon notification that they are loaded into memory. On reception of further notifications, if the checksum has changed, an alert is risen and further actions might be taken.
Let's now go into some details of the initialization, memorization and detection phase of the defensive driver. We'll end this paper by a demonstration of the defense granted by the modified driver against an attack by \Duqu.

\subsection{Initialization phase}

The initialization phase of our driver has been greatly simplified. We kept the creation of the access points, removed the search for the \texttt{ZwProtectVirtualMemory} function.
We kept the handling of notifications when modules are loaded and also asked for notifications when the system finishes creating a process (function \texttt{PsSetCreateProcessNotifyRoutine}).

\subsection{Memorization phase}
We saw that the alteration is not done when the first notification is triggered.
Thus, if a checksum had been done at that point, the modification of the entrypoint could be detected when the second notification is triggered.


We reused \Duqu's checksum function which was originally called for obfuscating function names.

A first notification is received when a process is created. Unfortunately Windows passes only the process \texttt{id}, its parent and whether it was created or destroyed. However we can retrieve the PEB (Process Enviroment Block) structure associated to that process. We used that information to find the memory address of the loaded file.
In order to detect \Duqu, we focused on \service, we compare the process name to the string "\service". If it is the target process, we look for its entrypoint, we calculate its hashed value and store it as an initial signature.
The defensive driver is now ready to detect the hook made by \Duqu.

\subsection{Detection phase}
When a module is loaded, the operating system passes control to the defensive driver which looks for the entrypoint of the module. If the loaded module is a DLL, the entrypoint is searched in the \texttt{PE} file of the executable file it is linked to.
We reused what was done in \Duqu\ and added the verification of the hash.



The hash of the PE file has been determined and is to be compared to the original signature previously stored.
If both checksums are different, we infer that \Duqu\ hooked \service\ between the two notifications. Since the entrypoint has been altered, the process \service\ is flagged as suspicious and submitted to further analysis.

\subsection{Proof of concept}
Aiming to debug and test the original driver and the defensive one, we followed the steps described by Sergei Shevchenko \cite{FSabatier_SShevchenko} who proposes to rename the Windows calculator \texttt{calc.exe} as \service\ and launch it to watch how the drivers react.

For this example we used two virtual machines using \texttt{Windows XPSP3} connected by a serial link. The first machine runs Microsoft's \texttt{WinDBG} for kernel debug. The other machine is launched in \texttt{"kernel debug"} mode which allows the debug machine to communicate on a kernel level and debug drivers.

While doing tests we uncovered that \Duqu's \texttt{nfrd965.sys} checks if the system is in debug or failsafe mode so we had to patch that out for test purposes, thus allowing debug.
We also configured both drivers so they can be launched on demand, it was needed to modify registry keys (the \texttt{Start} parameter has to be set to 3).
This configuration provides us with the possibility to choose the launch order of both drivers and \service.

We first launched the defensive driver, then \Duqu's driver and finally \service.
In the debugger console, shown on Figure  \ref{fig:AThierry_Breakpoint1}, we see when \service\ is launched. The system notifies the defensive driver which outputs information about the loaded module then stores its \texttt{id}, the address of its entrypoint and its initial signature (checksum of the first bytes at the entrypoint).

\begin{figure}

\scriptsize
\begin{lstlisting}[language={}]
-+* Create process 0x914 *+-
ProcessImageInformation: 
  PEB=0x7ffd6000 ImageBaseAddress=0x01000000
  UniqueProcessId=0x914 
Entrypoint bytes at 0x01012475: 
  0x6a 0x70 0x68 0xe0 0x15 0x00 0x01 0xe8
ProcessImageName: Desktop\services.exe
ProcessImageName: save processID=0x914
CreateProcessNotify: ImageBaseAddress=0x01000000
  EntryPoint=0x01012475
  EntrypointChecksum=0x49af1bf2
\end{lstlisting}
\caption{From WinDbg: The process services.exe is loaded. The defensive driver stores its id (\texttt{0x914}), its entrypoint (\texttt{0x01012475}) and its hash (\texttt{0x49af1bf2}).\label{fig:AThierry_Breakpoint1}}
\end{figure}

When the notification for \krn\ is triggered to the defensive driver, no modification has been made since \Duqu's driver will receive the notification afterwards (due to the launch order of drivers). So the checksum succeeds. However there are further notifications, for instance when the linked DLL \texttt{shell32.dll} is loaded, the defensive driver checks once again the entrypoint's hash and it has been altered. It is shown on Figure \ref{fig:AThierry_Breakpoint2}. Thus the alteration of \service\ is detected and the defensive driver takes further steps to protect the system: it terminates \service, ending \Duqu's attempt to compromise the machine.

\begin{figure}
\scriptsize
\begin{lstlisting}[language={}]
* Loaded module \WINDOWS\system32\kernel32.dll *
LoadImageNotifyRoutine:
  ImageBaseAddress=0x7c800000 ProcessId=0x914 
-> Verify services.exe process: 
   Entrypoint at 0x01012475:
     0x6a 0x70 0x68 0xe0 0x15 0x00 0x01 0xe8
-> OK!
* Loaded module \WINDOWS\system32\shell32.dll *
LoadImageNotifyRoutine:
  ImageBaseAddress=0x7c9d0000 ProcessId=0x914 
-> Verify services.exe process:
   Entrypoint at 0x01012475:
     0xb8 0xbd 0x18 0x0a 0x00 0xff 0xd0 0xe8
-> Checksum error !!!!
-> Terminating services.exe
\end{lstlisting}
\caption{Detection of the altered entrypoint (\texttt{0x01012475}) of services.exe.\label{fig:AThierry_Breakpoint2}}
\end{figure}

\section{Conclusion}
Similarities between \Duqu\ and \Stuxnet\ lead us to look for a detection method of \Duqu\ when the attack is going on.
We described by and large the infection technique of \Duqu\ and how its driver operates, stealthy injecting code into \service\ using kernel functions.
Thus we rebuilt a source code for \Duqu's driver and created a defensive version from this source code. Our modified driver is able to detect the injection made by \Duqu\ and protects the system by terminating the infected process \service.

\Duqu\ was considered at its times as one of the most sophisticated malware. And the above shows that, indeed, the malware was built with great care. At the same time, it is known that complex systems may be fragile. Usually, it is on the defender side that we make the observation: complex infrastructures offer a lot of entry points to malware. Here, the argument is opposite: it's the malware which was fragile and we exploited this feature to turn it for our own purposes.



%
~\\
\bibliographystyle{latex8} 
\bibliography{malware2013}

\begin{thebibliography}{1}\setlength{\itemsep}{-1ex}\small

\bibitem{AThierry_CVETrueType}
{CVE-2011-3402}.
\newblock
  \url{http://www.cve.mitre.org/cgi-bin/cvename.cgi?name=CVE-2011-3402}.

\bibitem{AThierry_IDADecompiler}
{Hex-Rays Decompiler}.
\newblock \url{http://www.hex-rays.com/products/decompiler/index.shtml}.

\bibitem{AThierry_BKM08}
G.~{B}onfante, M.~{K}aczmarek, and J.-Y. {M}arion.
\newblock {{A}rchitecture of a {M}orphological {M}alware {D}etector}.
\newblock {\em {J}ournal in {C}omputer {V}irology}, 5:263--270, 2009.

\bibitem{AThierry_CrysysDuquStuxnet}
L.~o. C. o. S.~S. (CrySyS).
\newblock {Duqu: A Stuxnet-like malware found in the wild}, October 2011.

\bibitem{AThierry_KaspDuqu10}
Kaspersky.
\newblock { The mystery of Duqu: Part Ten}, Mar. 2012.
\newblock
  \url{https://www.securelist.com/en/blog/208193425/The_mystery_of_Duqu_Part_Ten}.

\bibitem{FSabatier_SShevchenko}
S.~Shevchenko.
\newblock {Actually, my name is Duqu - Stuxnet is my middle name}.
\newblock
  \url{http://baesystemsdetica.blogspot.fr/2012/03/actually-my-name-is-duqu-stuxnet-is-my_4108.html}.

\bibitem{AThierry_SymantecDuqu2011}
Symantec.
\newblock {W32.Duqu: The Precursor to the Next Stuxnet}, October 2011.

\bibitem{AThierry_ThabetDriver}
A.~Thabet.
\newblock { Reversing Stuxnet's Rootkit (MRxNet) Into C++}, Jan. 2011.
\newblock
  \url{http://amrthabet.blogspot.fr/2011/01/reversing-stuxnets-rootkit-mrxnet-into.html}.

\bibitem{AThierry_REAT12}
A.~Thierry, G.~Bonfante, J.~Calvet, J.-Y. Marion, and F.~Sabatier.
\newblock {Recognition of binary patterns by Morphological analysis}.
\newblock {\em REcon}, 2012.

\end{thebibliography}

\end{document}